\documentclass[twocolumn,showpacs,preprintnumbers,amsmath,amssymb]{revtex4}
\usepackage{graphicx}
\usepackage{dcolumn}
\usepackage{bm}
\usepackage{hyperref}
\usepackage{xcolor}

\newcommand{\bef}{\begin{figure}}
\newcommand{\eef}{\end{figure}}

\newcommand{\be}{\begin{equation}}
\newcommand{\ee}{\end{equation}}
\newcommand{\bea}{\begin{eqnarray}}
\newcommand{\eea}{\end{eqnarray}}
\begin{document}

\title{Impact of particle production mechanisms on pseudorapidity distribution and directed flow in Au+Au and Cu+Cu collisions  at $\sqrt{s_{NN}}$ = 19.6 GeV using AMPT model}
\author{Muhammad Farhan Taseer and Subhash Singha}
\thanks{Contact Authors: \href{mailto:mfarhan\_taseer@impcas.ac.cn}{mfarhan\_taseer@impcas.ac.cn}, \href{mailto:subhash@impcas.ac.cn}{subhash@impcas.ac.cn}}
\affiliation{Institute of Modern Physics Chinese Academy of Sciences, Lanhzou 730000, China\\
University of Chinese Academy of Sciences, Beijing 100049, China}

\begin{abstract}

The STAR experiment at the top RHIC energy has observed that the directed flow ($v_1$) of inclusive light hadrons is independent of the collision system size at a given centrality~\cite{STAR:2008jgm}. However, recent STAR measurements indicate a system-size dependence in the $v_1(y)$-slope ($dv_{1}/dy$) of protons, antiprotons, and their differences ($\Delta dv_{1}/dy$) at a given centrality, suggesting a potential influence of baryon production and transport mechanisms~\cite{Taseer:SQM2024talk}. We have studied pseudorapidity ($dN/dy$) distributions and directed flow ($v_1$ and $dv_{1}/dy$) for pions, kaons, and protons in Au+Au and Cu+Cu collisions at $\sqrt{s_{NN}} = 19.6$ GeV using the A Multi-Phase Transport (AMPT) model. Specifically, we investigated the influence of string junction parameters in AMPT via the PYTHIA/JETSET routines, focusing on the popcorn mechanism and string-splitting parameters, on $dN/dy$, $dv_{1}/dy$, and their charge-dependent splittings ($\Delta dN/dy$ and $\Delta dv_{1}/dy$). We observe that string junction parameters can affect $dN/dy$, $dv_{1}/dy$, $\Delta dN/dy$, and $\Delta dv_{1}/dy$ for $\pi$, K, and p, and influence their system-size dependence. The effect is most prominent on the $v_1$ of protons, non-trivial for kaons, while the pions $v_1$ remain largely unchanged. These findings provide insights into the interplay between particle production mechanisms, baryon transport, and directed flow in heavy-ion collisions.
 
\end{abstract}
\pacs{25.75.Ld}
\maketitle

\section{INTRODUCTION}
Relativistic heavy-ion collisions are performed to study the behavior of a strongly interacting matter comprising quarks and gluons, called Quark Gluon Plasma (QGP) governed by Quantum Chromodynamics (QCD). From the heavy-ion collision experiments at facilities, such as RHIC and LHC~\cite{Adams:2005dq,Adcox:2004mh,BRAHMS:2004adc,Harris:1996zx,Muller:2012zq,Braun-Munzinger:2007edi}, an existance of a strongly coupled QGP is established. Measurement of collective flow, described by the Fourier coefficients of azimuthal angle of the produced particles, has widely used to study the properties and dynamics of QGP~\cite{Ollitrault:1992bk, Poskanzer:1998yz, Voloshin:1994mz}. The first Fourier coefficient of anisotropic flow, called directed flow ($v_{1}$)~\cite{Brachmann:1999xt}, is of special interest because it is sensitive to the early time dynamics and equation of state of the systems. Typically the $v_{1}$ of produced particles is measured following 

\begin{equation}
v_{1} = \langle cos (\phi - \Psi_{RP}) \rangle, 
\end{equation}
where $\phi$ is the azimuthal angle and $\Psi_{RP}$ is the reaction plane subtended by the impact parameter and the beam-axis. The average $\langle ... \rangle$ is done over all particles and all events. Primarily we focus on the rapidity-odd component of directed flow. Extensive research on $v_{1}$ of several identified particles has been conducted at AGS~\cite{E895:2000sor, Chung:2001je}, SPS~\cite{NA49:2003njx}, RHIC~\cite{STAR:2003xyj, PHOBOS:2005ylx}, and LHC~\cite{ALICE:2013xri} energies. One of the key interests has been to search for a change in the sign of proton and net-proton $v_{1}$, due to its potential connection with the softening of the Equation of State (EoS) associated with the QCD phase transition~\cite{Stoecker:2004qu, Nara:2016phs, Konchakovski:2014gda}. The STAR experiment has observed a sign change in proton $v_1$ and a double sign change in net-proton $v_1$ near $\sqrt{s_{NN}} = 10$–$20$ GeV~\cite{STAR:2014clz, STAR:2017okv}. However, to date, no model has been able to accurately capture this pattern or the position of the sign change, making $v_1$ one of the most challenging observables to model~\cite{Singha:2016mna}.

Furthermore, the rapidity-odd $v_{1}$ is also sensitive to the electromagnetic field effects in heavy-ion collisions~\cite{Kharzeev:2007jp, McLerran:2013hla} and usually probed via the difference in $v_{1}$ of particles and anti-particles (called $\Delta v_1$)~\cite{Gursoy:2014aka,Gursoy:2018yai}. Recent high-statistics data from the STAR experiment revealed intriguing features in $\Delta v_1$ for light-quark hadrons. At top RHIC energy, $\Delta v_1$ is positive in (mid-)central collisions for protons and kaons, but becomes significantly negative in peripheral collisions. The positive values are attributed to transported quarks, while the negative values are consistent with electromagnetic effects dominated by Faraday induction~\cite{STAR:2023jdd}. Additionally, $\Delta v_1$ was found to increase with both electric charge and strangeness differences in mid-central collisions~\cite{STAR:2023wjl}. More recently, STAR reported simultaneous measurements of $v_1$ and $\Delta v_1$ for $\pi$, $K$, and $p$ in $U+U$, $Au+Au$, and isobar ($Ru+Ru$, $Zr+Zr$) collisions~\cite{Taseer:SQM2024talk}. A clear system-size dependence of $v_1$ and $\Delta v_1$ is observed for protons and antiprotons, with ordering $\Delta v_1$: $U+U > Au+Au >$ ($Ru+Ru$, $Zr+Zr$), especially in central and mid-central collisions, while mesons show no such dependence. This apparently contrasts earlier RHIC results of inclusive charged particles where $v_1$ is appeared system-size independent. A relativistic dissipative hydrodynamic model with baryon diffusion~\cite{Parida:2023ldu, Parida:2023rux} qualitatively reproduces the observed $\Delta v_1$ trends for baryons and their system-size dependence, but fails to simultaneously describe the $\Delta v_1$ sign and centrality trends across $\pi^{\pm}$, $K^{\pm}$, and $p (\bar{p})$~\cite{Parida:2025ddt}. At lower energies, mean-field potentials may significantly influence $v_1$ and its charge dependence~\cite{Isse:2005nk, Nara:2021fuu, Nara:2020ztb}, though these effects are not considered in the present study.

This paper aims to investigate the directed flow ($v_1$) and its system-size dependence for $\pi$, $K$, and $p$ at lower beam energies, where the magnitude of $v_1$ is significantly larger, making it easier to analyze. For this purpose, we utilize Au+Au and Cu+Cu collisions at $\sqrt{s_{NN}} = 19.6$ GeV within the A Multi-Phase Transport (AMPT) model framework~\cite{Lin:2004en}. To gain insights into the connection between the particle production mechanisms near mid-rapidity and their influence on directed flow, we also analyze the pseudorapidity distributions of $\pi$, $K$, and $p$, along with their charge-dependent differences.

\begin{figure*}
\begin{center}
\includegraphics[scale=0.7]{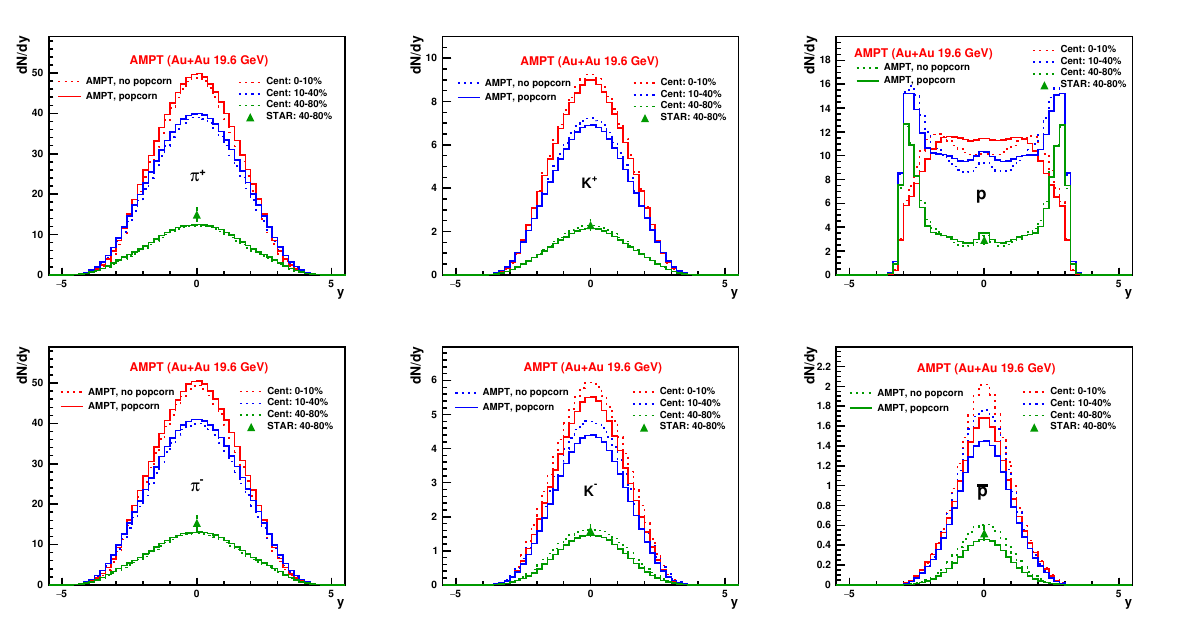}
\caption{Pseudorapidity ($dN/dy$) distribution for $\pi$, $K$, and $p$ and their antiparticles in Au+Au collisions at $\sqrt{s_{NN}}$ = 19.6 GeV for different collision centralities with and without popcorn mechanism (Table I: set-1 and set-3) using AMPT model. The STAR data points ($|y|<0.1$) are taken from~\cite{STAR:2017sal}}
\label{fig1_dndyauau}
\end{center}
\end{figure*} 

\begin{figure*}
\begin{center}
\includegraphics[scale=0.7]{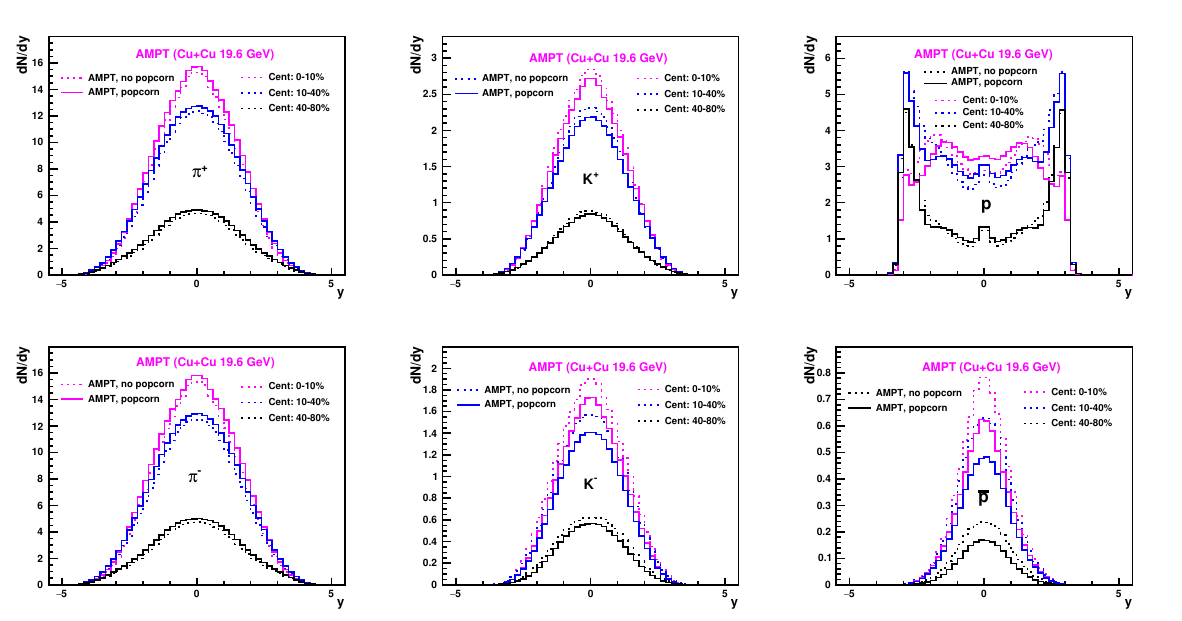}
\caption{Pseudorapidity ($dN/dy$) distribution for $\pi$, $K$, and $p$ and their antiparticles in Cu+Cu collisions at $\sqrt{s_{NN}}$ = 19.6 GeV for different collision centralities with and without popcorn mechanism (Table I: set-1 and set-3) using AMPT model.}
\label{fig2_dndycucu}
\end{center}
\end{figure*} 

\begin{figure*}
\begin{center}
\includegraphics[scale=0.7]{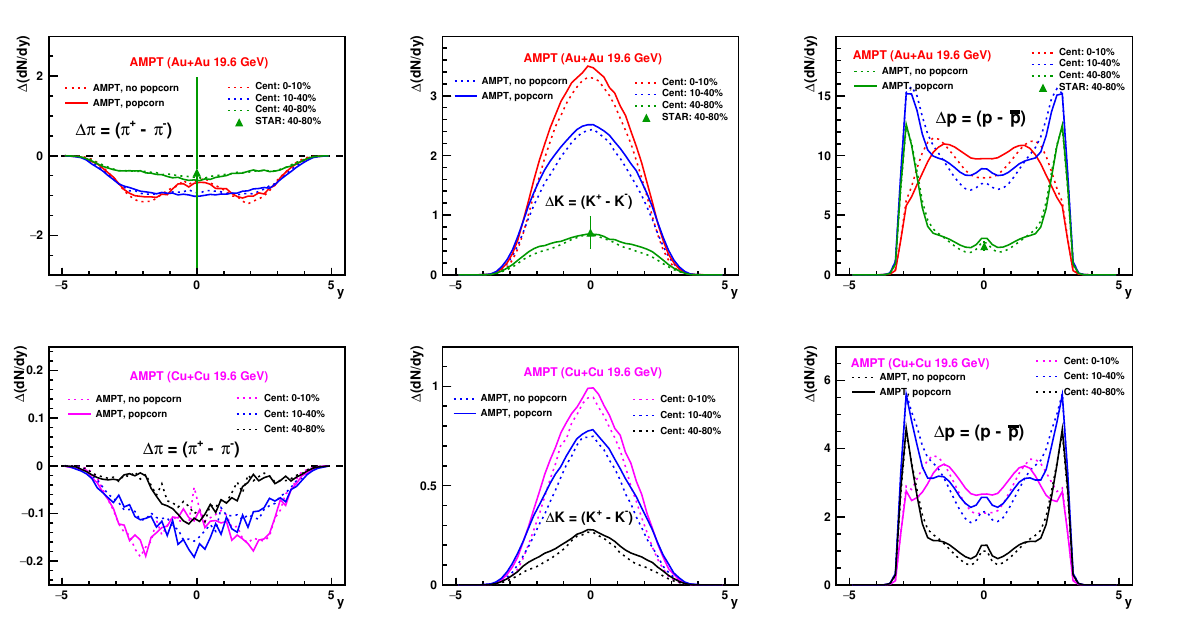}
\caption{Pseudorapidity ($dN/dy$) distribution for net-$\pi$, net-$K$, and net-$p$ in Au+Au and Cu+Cu collisions at $\sqrt{s_{NN}}$ = 19.6 GeV for different collision centralities with and without popcorn mechanism (Table I: set-1 and set-3) using AMPT model. The STAR data points ($|y|<0.1$) are taken from~\cite{STAR:2017sal}}
\label{fig3_deltadndyauaucucu}
\end{center}
\end{figure*} 

\begin{figure*}
\begin{center}
\includegraphics[scale=0.7]{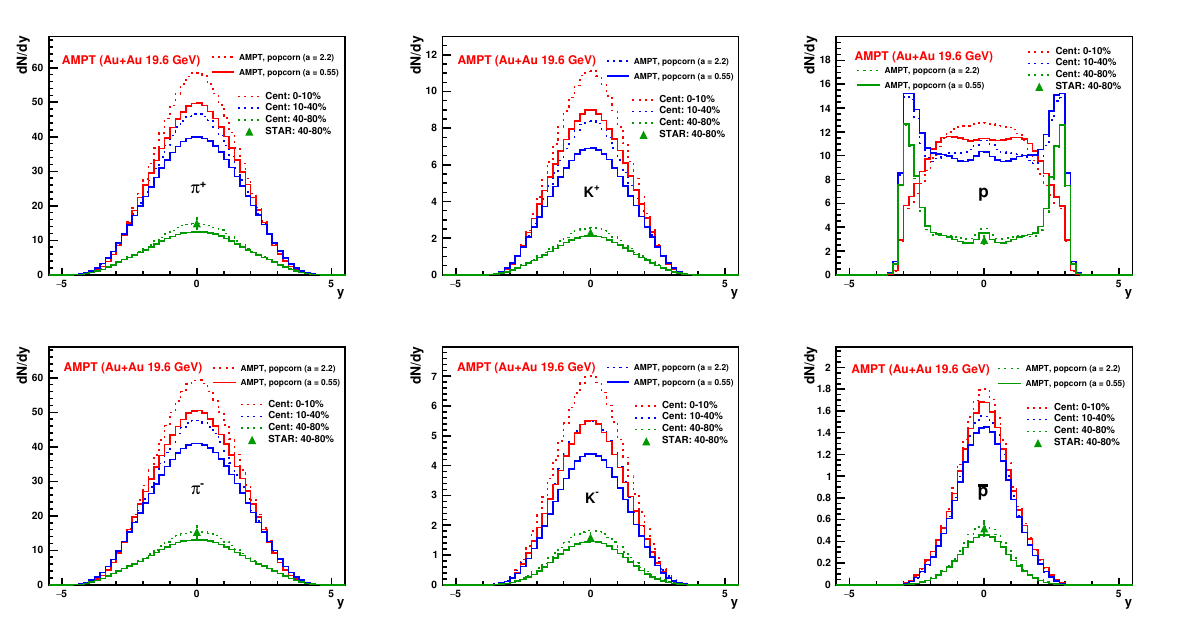}
\caption{Pseudorapidity ($dN/dy$) distribution for $\pi$, $K$, and $p$ in Au+Au collisions at $\sqrt{s_{NN}}$ = 19.6 GeV for different collision centralities with two cases of popcorn mechanism, Lund fragmentation parameter a=0.55 and a=2.2 (Table I: set-3 and set-4), using AMPT model. The STAR data points ($|y|<0.1$) are taken from~\cite{STAR:2017sal}}
\label{fig4_dndy_apara_auau}
\end{center}
\end{figure*} 

%
\begin{figure*}
\begin{center}
\includegraphics[scale=0.7]{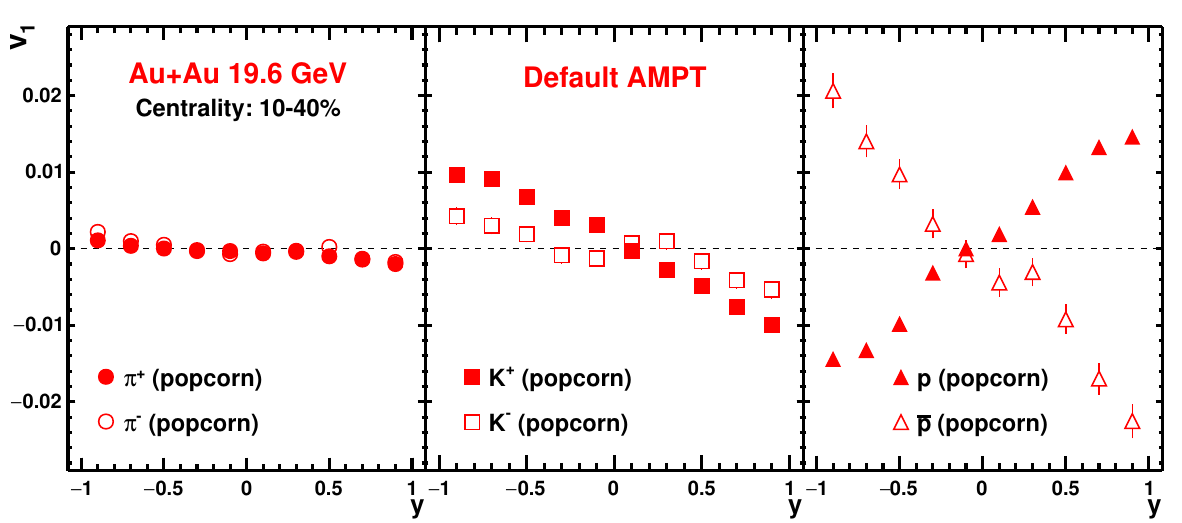}
\includegraphics[scale=0.7]{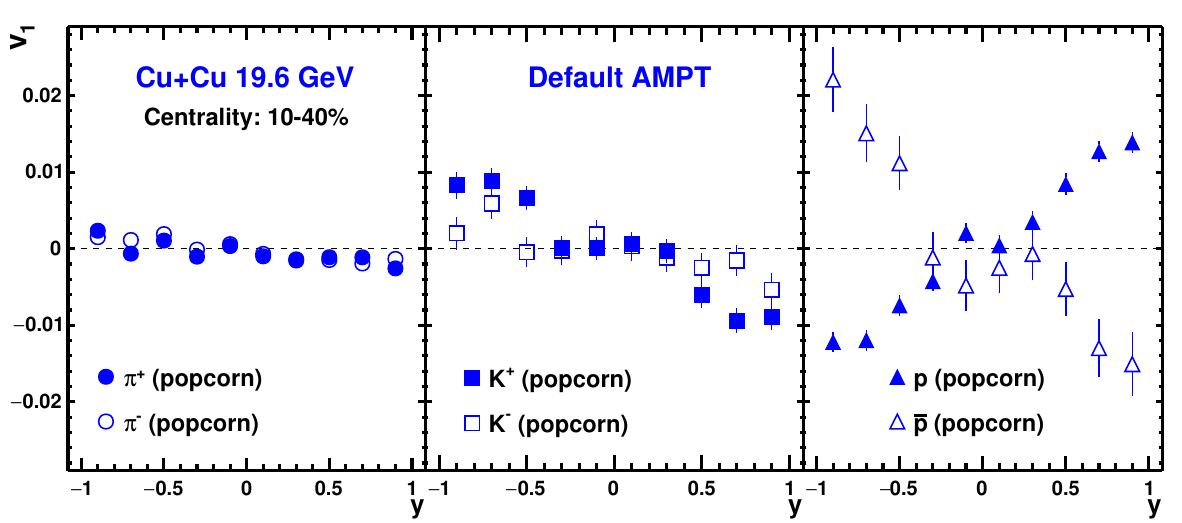}
\caption{Directed flow $v_{1} (y)$ for $\pi^{+}$, $K^{+}$, $p$, and their antiparticles in Au+Au and Cu+Cu collisions at $\sqrt{s_{NN}}$ = 19.6 GeV for 10-40\% centrality with popcorn mechanism and with Lund fragmentation parameter a=0.55, b=0.15 GeV$^{2}$ (Table I: set-3) using AMPT model.}
\label{fig5ab_v1y_auau_cucu}
\end{center}
\end{figure*} 

\begin{figure*}
\begin{center}
\includegraphics[scale=0.7]{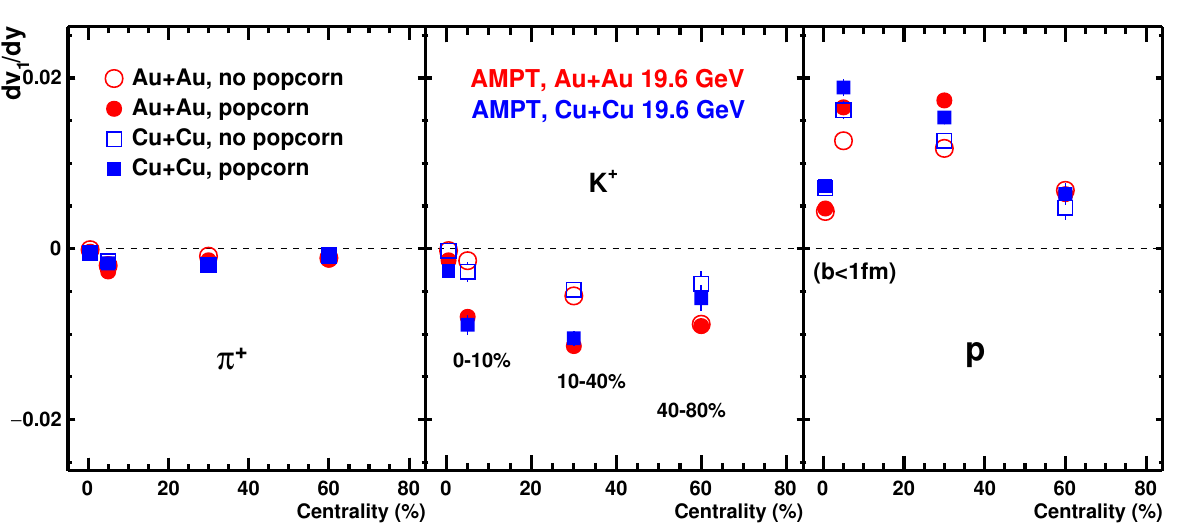}
\includegraphics[scale=0.7]{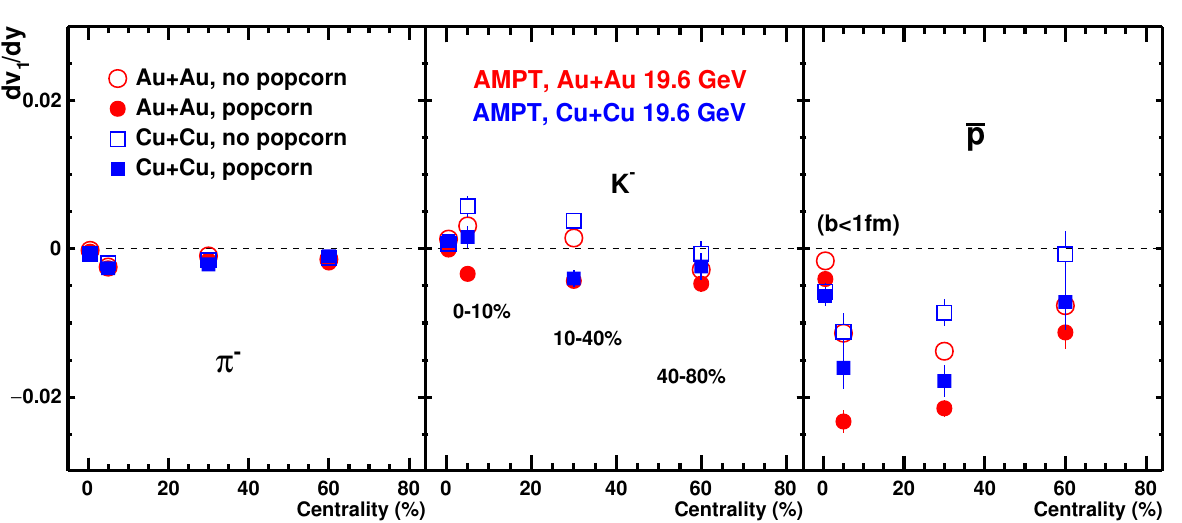}
\caption{$dv_{1}/dy$ for $\pi^{+}$, $K^{+}$, and $p$ and their antiparticles in Au+Au and Cu+Cu collisions at $\sqrt{s_{NN}}$ = 19.6 GeV for different collision centralities with and without popcorn mechanism (Table I: set-1 and set-3) using AMPT model. The model points at centrality = 0.5 corresponds to ultra-central events ($b<1$ fm/c).}
\label{fig6_dv1dy_w_wo_pop_auaucucu}
\end{center}
\end{figure*} 

\begin{figure*}
\begin{center}
\includegraphics[scale=0.7]{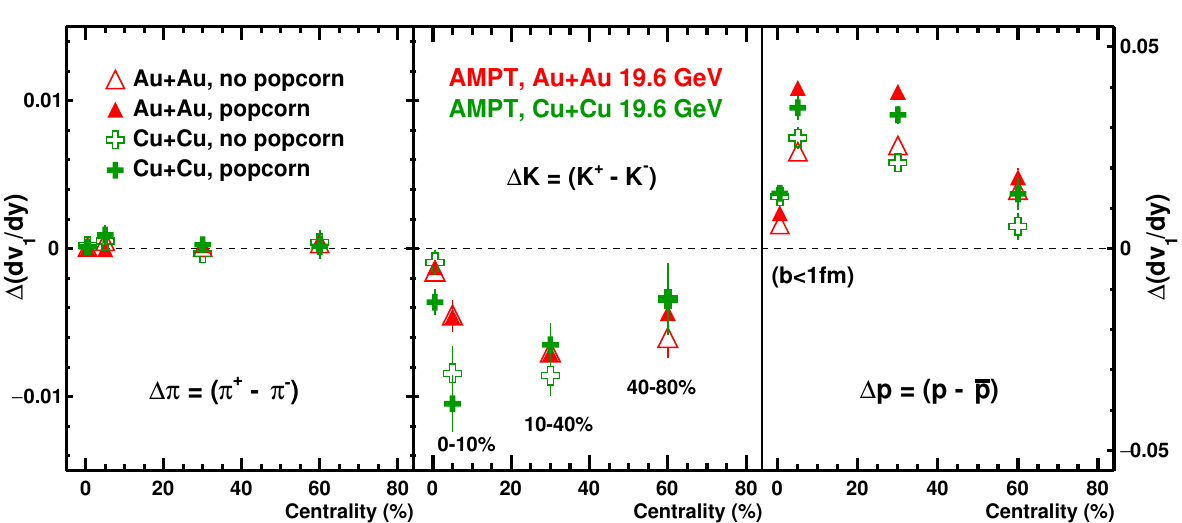}
\caption{$\Delta dv_{1}/dy$ for $\pi$, $K$, and $p$ in Au+Au and Cu+Cu collisions at $\sqrt{s_{NN}}$ = 19.6 GeV for different collision centralities with and without popcorn mechanism (Table I: set-1 and set-3) using AMPT model. The model points at centrality = 0.5 corresponds to ultra-central events ($b<1$ fm/c).}
\label{fig7_deldv1dy_w_wo_pop_auaucucu}
\end{center}
\end{figure*} 

\begin{figure*}
\begin{center}
\includegraphics[scale=0.7]{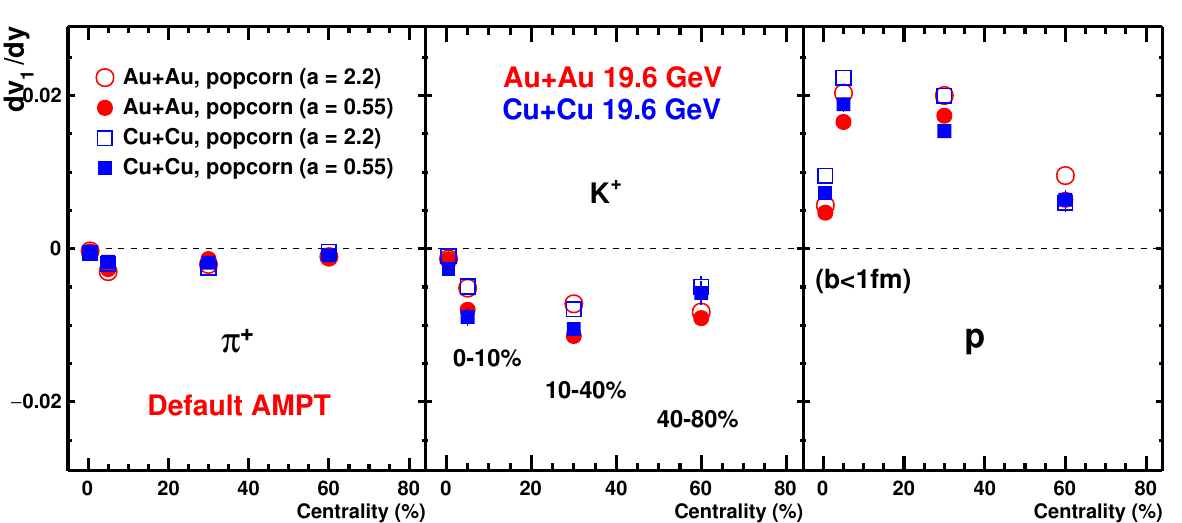}
\includegraphics[scale=0.7]{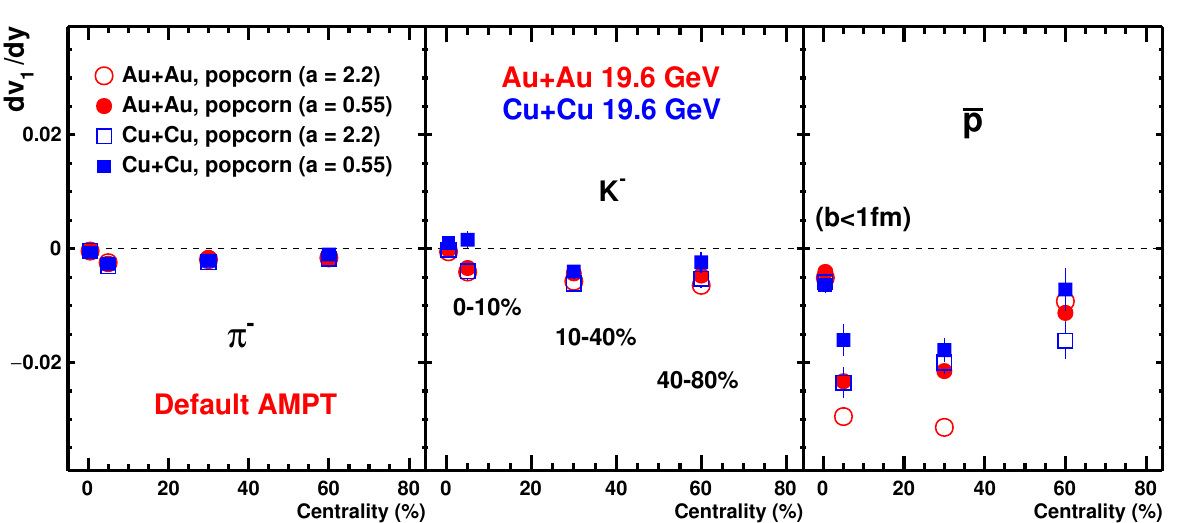}
\caption{$dv_{1}/dy$ for $\pi^{+}$, $K^{+}$, and $p$ and their antiparticles in Au+Au and Cu+Cu collisions at $\sqrt{s_{NN}}$ = 19.6 GeV for different collision centralities with two cases of popcorn mechanism, Lund fragmentation parameter a=0.55 and a=2.2 (Table I: set-3 and set-4), using AMPT model. The model points at centrality = 0.5 corresponds to ultra-central events ($b<1$ fm/c).}
\label{fig8_dv1dy_a_par_auau}
\end{center}
\end{figure*} 

\begin{figure*}
\begin{center}
\includegraphics[scale=0.7]{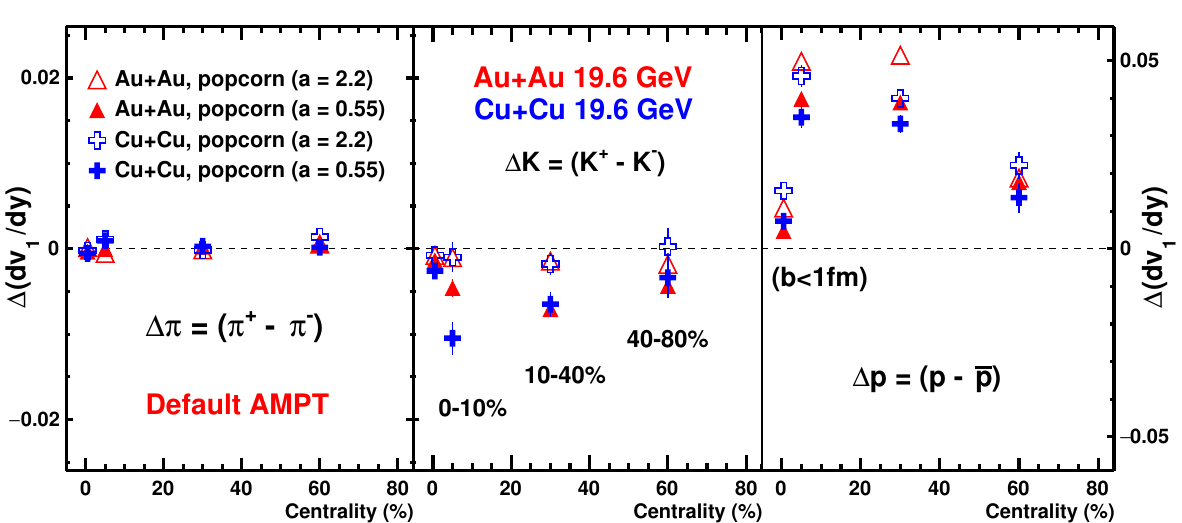}
\caption{$\Delta dv_{1}/dy$ for $\pi$, $K$, and $p$ in Au+Au and Cu+Cu collisions at $\sqrt{s_{NN}}$ = 19.6 GeV for different collision centralities with two cases of popcorn mechanism, Lund fragmentation parameter a=0.55 and a=2.2 (Table I: set-3 and set-4), using AMPT model. The model points at centrality = 0.5 corresponds to ultra-central events ($b<1$ fm/c)}
\label{fig9_dv1dy_a_par_auau}
\end{center}
\end{figure*} 

\begin{figure*}
\begin{center}
\includegraphics[scale=0.7]{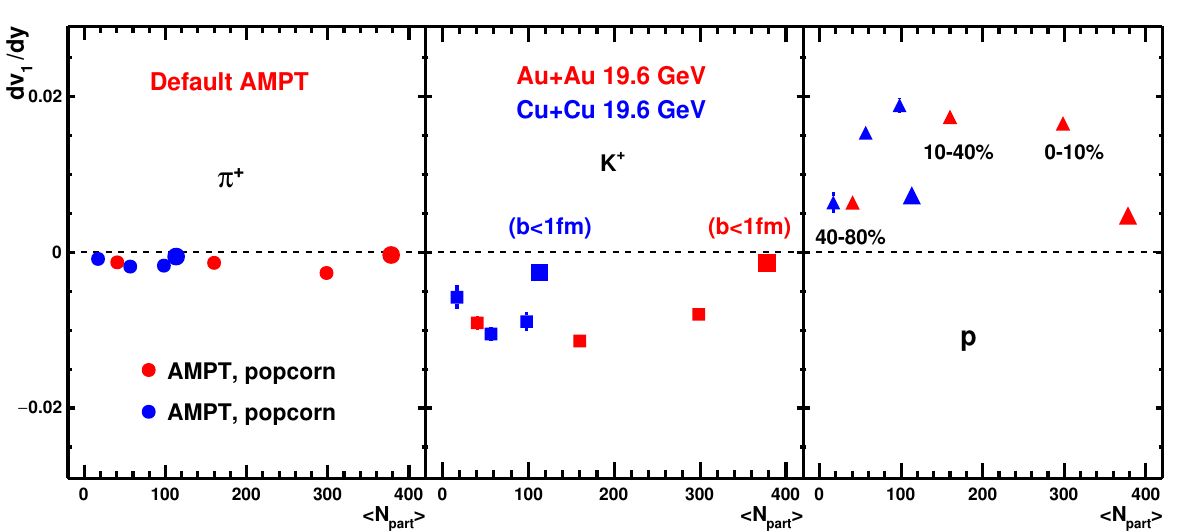}
\includegraphics[scale=0.7]{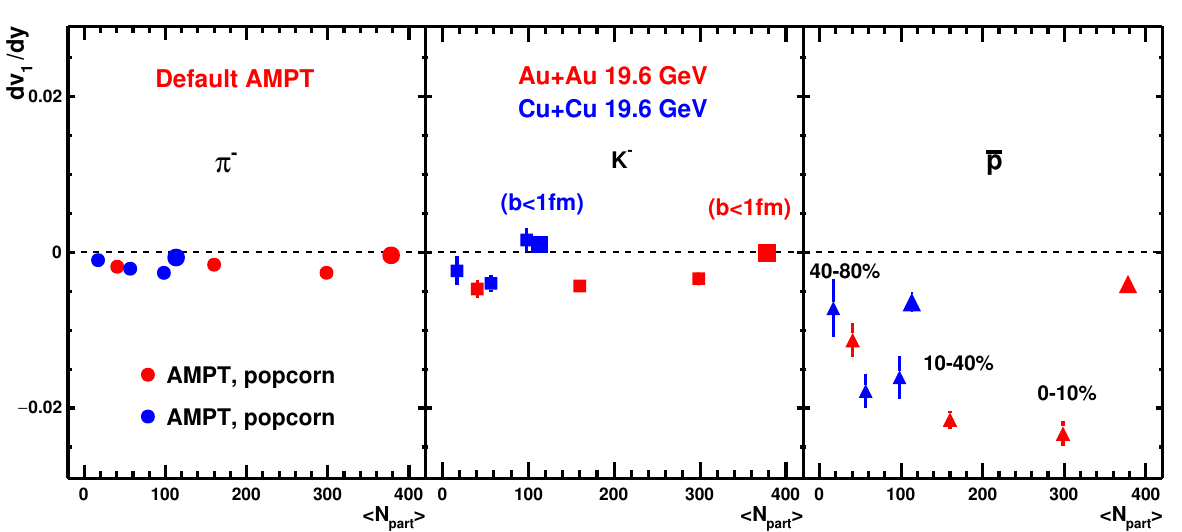}
\caption{$dv_{1}/dy$ for $\pi$, $K$, and $p$ and their antiparticles in Au+Au and Cu+Cu collisions as a function of average number of participants ($\langle N_{part} \rangle$) at $\sqrt{s_{NN}}$ = 19.6 GeV  with popcorn mechanism (Table I: set-3) using AMPT model. The model points with larger marker size corresponds to ultra-central events ($b<1$ fm/c).}
\label{fig10_dv1dy_npart_auau_cucu}
\end{center}
\end{figure*} 

\begin{figure*}
\begin{center}
\includegraphics[scale=0.7]{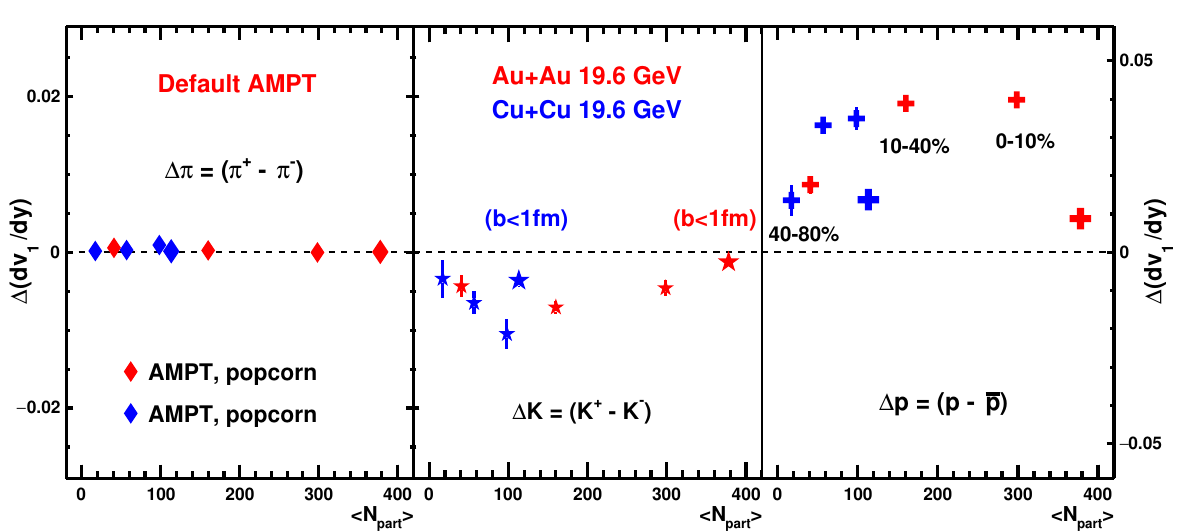}
\caption{$\Delta dv_{1}/dy$ for $\pi$, $K$, and $p$ in Au+Au and Cu+Cu collisions as a function of average number of participants ($\langle N_{part} \rangle$) at $\sqrt{s_{NN}}$ = 19.6 GeV  with popcorn mechanism (Table I: set-3) using AMPT model. The model points with larger marker size corresponds to ultra-central events ($b<1$ fm/c).}
\label{fig11_dv1dy_npart_auau_cucu}
\end{center}
\end{figure*} 

\section{The AMPT Model}

In this analysis we have utilized the default version of AMPT model~\cite{Lin:2004en}, which is a hybrid transport model consisting of four major components: (i) The initial conditions are obtained from HIJING~\cite{Wang:1991hta}. Minijets remain coexists with the remnants of their originating nucleons, collectively forming newly excited strings; (ii) The scattering among partons are modeled by Zhang's Parton Cascade (ZPC)~\cite{Zhang:1997ej}; (iii) When the minijet partons stop interacting, they are combined with their parent strings to form excited strings, which are then converted to hadrons according to the Lund string fragmentation model~\cite{Andersson:1983jt, Andersson:1983ia, Sjostrand:1993yb}; (iv) finally, the evolution of hadronic matter is described by a hadron cascade which built upon A Relativistic Transport (ART) framework~\cite{Li:1995pra}. The ART accounts for both elastic and inelastic scatterings involving baryon-baryon, baryon-meson, and meson-meson interactions. 

In the string fragmentation model, mesons are conceptualized as having a short string connecting their quark and antiquark endpoints, whereas baryons are described as three-quark structures. The PYTHIA/JETSET subroutines within the AMPT model incorporate particle production based on Lund string fragmentation model~\cite{Sjostrand:1993yb, Lin:2004en}. The probability of a string splitting into hadrons is governed by Lund symmetric fragmentation function:
\begin{equation}
f(z) \propto z^{-1} \; (1-z)^{a} \; exp(-bm_{T}^{2}/z)
\end{equation}
where $z$ correspond to the fraction of string’s energy-momentum taken by one fragment; $m_{T} = \sqrt(m^2 + p_{T}^{2})$ transverse mass of the produced hadrons; $a$ and $b$  parameters controlling the shape of the splitting function; and $m$ is the mass of the produced hadron. These two parameters $a$ and $b$ in the Lund fragmentation function are approximately related to the string tension ($\kappa$) by,
\begin{equation}
\kappa \propto \frac{1}{b(2+a)}
\end{equation}
The value of $a$ and $b$ parameters can be controlled by \texttt{PARJ(41)} and \texttt{PARJ(42)} parameters in AMPT. The $a$ and $b$ were determined by fitting the charged particle yield and $p_{T}$ spectra in different collision systems. Note that a larger value of $a$ corresponds to a softer fragmentation function, leading to a smaller average transverse momentum for produced hadrons and thus increases the particle multiplicity. The default values used in AMPT are $a=0.55$ and $b=0.15$ GeV$^{-2}$. We have varied $a=2.2$, $b=0.15$ GeV$^{-2}$ to check its impact on $dN/dy$ and especially on $v_{1}$ of identified particles. 

Furthermore, in string breaking picture a straightforward approach involves the diquark model, where quarks are treated as either individual quarks or antidiquarks in a color triplet state. In this framework, baryons and antibaryons are formed as neighboring entities along the string. An alternative mechanism, referred to as the "popcorn mechanism", allows for baryon production without direct diquark formation. Instead, baryons emerge through the sequential production of quark-antiquark pairs facilitated by fluctuations in the color field. The basic version of this mechanism permits at most one intermediate meson to form during the process (e.g., a $q\bar{q}$ string breaking into a $BM\bar{B}$ structure). An extended version of this mechanism, however, allows for the creation of multiple mesons during the string breaking process (e.g., a $q\bar{q}$ string breaking into a $BMM\bar{B}$ structure), as detailed in related studies. These string breaking processes can significantly influence meson and baryon production near mid-rapidity and has been studied using AMPT model~\cite{Lin:2000cx}. The stopping of baryons and their transport to mid-rapidity is a key phenomenon extensively studied in high-energy nuclear collisions. This process is closely related to the total energy deposited in the system. The observation of increase in stopped baryons near mid-rapidity has led to the proposal of novel mechanisms, such as the gluon junction mechanism. In the AMPT model, baryon stopping is phenomenologically incorporated via the "popcorn" mechanism. The parameter \texttt{MSTJ(12)} governs the operation of the popcorn mechanism. When \texttt{MSTJ(12) = 0}, the string breaks directly into a baryon-antibaryon pair without the production of intermediate mesons. Whereas, values of \texttt{MSTJ(12) > 0} enable the creation of intermediate mesons. Another parameter, \texttt{PARJ(5)}, determines the probability of meson production between baryon-antibaryon pairs when the popcorn mechanism is active. In PYTHIA, the \texttt{PARJ(5)} parameter is related to baryon and meson production via relation
\begin{equation}
\frac{P(BM\bar{B})}{(P(B\bar{B})+P(BM\bar{B}))} = \frac{\texttt{PARJ(5)}}{(0.5+\texttt{PARJ(5)})}
\end{equation}
This parameter controls the relative percentage of $B\bar{B}$ and $BM\bar{B}$ channels. We have studied the impact of \texttt{MSTJ(12)} and \texttt{PARJ(5)} on the $dN/dy$ and $v_{1}$ of identified particles. 

\begin{table}
\caption{The Lund string fragmentation parameters used in this AMPT simulation.}
\begin{tabular}{|c|c|c|c|c|}
\hline
Parameters &  \texttt{MSTJ(12)} &  \texttt{PARJ(5)} &  \texttt{PARJ(41)} & \texttt{PARJ(42)}\\ 
\hline
set-1: no-popcorn & 0 & 0 & 0.55 & 0.15\\
\hline
set-2: popcorn & 1 & 0.5 & 0.55 & 0.15\\
\hline
set-3: popcorn & 1 & 1 & 0.55 & 0.15\\
\hline
set-4: popcorn & 1 & 1 & 2.2 & 0.15\\
\hline
\end{tabular}
\label{tab1_settings}
\end{table}

\section{Results and Discussion}

We perform AMPT simulations comprising approximately 1 million events for each of the previously mentioned configurations, as defined in Table~\ref{tab1_settings}, covering an impact parameter range from 0.0 to 15.0 fm. Figures~\ref{fig1_dndyauau} and \ref{fig2_dndycucu} show the pseudorapidity distribution ($dN/dy$) for $\pi^{+}$, $K^{+}$, and $p$ and their antiparticles, respectively, in Au+Au and Cu+Cu collisions at $\sqrt{s_{NN}}$ = 19.6 GeV for three different collision centralities, with and without the popcorn mechanism. Switching on the popcorn mechanism in AMPT with \texttt{PARJ(5)}=1 increases the yields for $\pi^{+}$ ($\pi^{-}$) and $p$ near mid-rapidity, while it decreases the $K^{\pm}$ and $\bar{p}$ yield. A similar feature is observed for both Au+Au and Cu+Cu collisions. For the net particle pseudorapidity distribution (denoted $\Delta dN/dy$, Fig.~\ref{fig3_deltadndyauaucucu}), defined as the difference between particle and antiparticle yields, the popcorn mechanism increases the mid-rapidity yield for kaons and protons, while for pions, the relative change is small. In previous AMPT study,  it is observed that with equal probabilities for the $B\bar{B}$ and $BM\bar{B}$ configurations can reproduce the net-baryon rapidity distribution at SPS energies. It is also observed that without the popcorn mechanism, as implemented in default HIJING model, the net-baryon rapidity distribution would peak at a larger rapidity. Fig.~\ref{fig4_dndy_apara_auau} presents the $dN/dy$ distribution for two different implementations of the Lund string parameter \(a\), with the popcorn mechanism: \(a = 0.55\) and \(a = 2.2\). As expected, larger values of \(a\) increase the mid-rapidity yields for all species (\(\pi\), \(K\), and \(p\)). By varying the Lund string fragmentation mechanism, we altered the mid-rapidity yields of particles. Next, we studied the impact of these changes on the directed flow. As discussed in~\cite{Lin:2004en}, the default $a$ and $b$ parameters in HIJING can reproduce charged particle multiplicities in elementary collisions. Consequently, these parameters in the AMPT model may better describe peripheral collisions compared to central collisions. Although to better explain the data, the $a$ and $b$ values in the AMPT model can be varied with centrality, we have not yet investigated the extent of this variation. We have also compared our calculations of mid-rapidity $dN/dy$ and $\Delta dN/dy$ with STAR data ($|y|<0.1$) for Au+Au collisions~\cite{STAR:2017sal}, and observed that the results from the 40–80\% centrality bins are comparable with AMPT, whereas AMPT underpredicts the values for mid-central and central collisions.

The top and bottom panels in Fig.~\ref{fig5ab_v1y_auau_cucu} presents the rapidity dependence of $v_{1}$ for $\pi^{+}$, $K^{+}$, $p$, and their antiparticles  for mid-central (10-40\%) Au+Au and Cu+Cu collisions respectively at $\sqrt{s_{{NN}}} = 19.6$ GeV using popcorn process in AMPT. The $v_{1} (y)$ is fitted with a linear function within $-1.0 < y < 1.0$ to extract the slope near mid-rapidity or $dv_{1}/dy$. The $dv_{1}/dy$ is studied as function of centrality and also for ultra-central collisions ($b<1$ fm/c).

Fig.~\ref{fig6_dv1dy_w_wo_pop_auaucucu} shows $dv_{1}/dy$ as a function of centrality for $\pi^{+}$, $K^{+}$, $p$, and their antiparticles in Au+Au and Cu+Cu collisions at $\sqrt{s_{{NN}}} = 19.6$ GeV. Results with and without the popcorn mechanism are shown. It is observed that the popcorn process increases the magnitude of $dv_{1}/dy$ for $K^{+}$, $p$, and $\bar{p}$, while the effect on pions is small. In contrast, for 0-10\%  and 10-40\% collisions, a sign reversal is observed in $dv_{1}/dy$ for $K^{-}$, when the popcorn mechanism is enabled. Comparing the smaller Cu+Cu system to the larger Au+Au system, pions and, to a large extent, kaons do not exhibit substantial changes in $dv_{1}/dy$, whereas protons (and anti-protons) show a more pronounced effect. For ultra-central collisions, the directed flow is expected to vanish due to the absence of longitudinal asymmetry. We observed that for pions and kaons, $dv_{1}/dy$ is nearly zero for $b<1$ fm/c, whereas protons and antiprotons exhibit a considerable magnitude. The non-zero magnitude of $p$
($\bar{p}$) may be related to contributions from transported baryons or event-by-event asymmetries in baryon transport and requires further investigation.

Fig.~\ref{fig7_deldv1dy_w_wo_pop_auaucucu} presents the difference in $dv_{1}/dy$ for positive and negative particles, referred to as $\Delta dv_{1}/dy$, as a function of centrality for $\pi$, $K$, and $p$ in Au+Au and Cu+Cu collisions. The $\Delta dv_{1}/dy$ for pions is small in magnitude, while for kaons, it is negative. For protons, however, $\Delta dv_{1}/dy$ is positive and exhibits a clear centrality dependence, with a maximum observed in mid-central (10–40\%) collisions. In the same 10–40\% centrality range, $\Delta dv_{1}/dy$ for protons is larger in Au+Au collisions compared to Cu+Cu collisions. In contrast, $\Delta dv_{1}/dy$ for pions do not exhibit system-size dependence. For kaons, an opposite system size ordering to that of protons is observed: the magnitude of $\Delta dv_{1}/dy$ in Cu+Cu is larger than in Au+Au, especially in central collisions, while in peripheral collisions no obvious system size dependence is seen.

Next, we examine the effect on $dv_{1}/dy$ and $\Delta dv_{1}/dy$ for the aforementioned particles by varying the Lund string fragmentation parameter $a$ in both Au+Au and Cu+Cu collisions, presented in Fig.~\ref{fig8_dv1dy_a_par_auau}. As the value of $a$ increases from 0.55 to 2.2, the magnitude of $dv_{1}/dy$ increases for $K^{-}$, $p$, and $\bar{p}$, whereas $K^{+}$ shows a decreasing trend. In contrast, $\pi^{\pm}$ remain unaffected by this variation. The effect of the parameter $a$ on $\Delta dv_{1}/dy$ is shown in Figure~\ref{fig9_dv1dy_a_par_auau}. With the increase in $a$, the magnitude of $\Delta dv_{1}/dy$ decreases for kaons increases for protons, and remains largely unchanged for pions.

Furthermore, we analyzed the variation of $dv_{1}/dy$ and $\Delta dv_{1}/dy$ in Au+Au and Cu+Cu collisions as a function of the average number of participating nucleons ($\langle N_{part} \rangle$), as shown in Figs.~\ref{fig10_dv1dy_npart_auau_cucu} and \ref{fig11_dv1dy_npart_auau_cucu}. At small $\langle N_{part} \rangle$ values, $dv_{1}/dy$ and $\Delta dv_{1}/dy$ for all particles approximately scale with $\langle N_{part} \rangle$, but this scaling breaks down at large $\langle N_{part} \rangle$.

\section{Summary and Conclusion}
In summary, we present measurements of pseudorapidity distributions ($dN/dy$) and directed flow ($v_1$) for pions, kaons, and protons in Au+Au and Cu+Cu collisions at $\sqrt{s_{\mathrm{NN}}} = 19.6$ GeV, using the AMPT model. The string junction parameters in AMPT, specifically the popcorn mechanism and the string splitting parameter, are tuned to modify the production and yield of pions, kaons, and protons near mid-rapidity. By utilizing the string fragmentation mechanism and varying its parameters, we observe changes in the fraction of net particles near mid-rapidity.

When the popcorn mechanism is enabled, we observe an increase in both $dN/dy$ and the magnitude of $dv_{1}/dy$ for protons, as well as in $\Delta dN/dy$ and $\Delta dv_{1}/dy$ for protons ($\Delta p$), within a given collision system such as Au+Au or Cu+Cu. For pions, $dN/dy$ increases, but no appreciable change is observed in $dv_{1}/dy$ or $\Delta dv_{1}/dy$. In contrast, kaons exhibit non-trivial changes in both yield and $dv_{1}/dy$.

When the string-splitting parameter $a$ is increased from 0.5 to 2.2, the rapidity density $dN/dy$ increases for all particle species. This change in $a$ also enhances the magnitude of $dv_{1}/dy$ for $p$, $\bar{p}$, and $\Delta p$. However, $dv_{1}/dy$ for $K^{+}$ and $\Delta K$ exhibits the opposite trend, with a decrease in magnitude, while pions remain largely unaffected.

Additionally, comparisons between different collision systems, such as Au+Au and Cu+Cu, reveal that both mechanisms, namely turning the popcorn process and varying the string splitting parameter, can contribute to the system size dependence of $dv_{1}/dy$ for protons at fixed centrality, along with a non-trivial effect on kaons, while pions remain largely unchanged across system sizes. These findings can provide valuable insights into the particle production mechanisms and directed flow across different system size.

In the future, these studies can be extended to both higher and lower beam energies. Notably, none of the current models can simultaneously capture the negative $\Delta dv_{1}/dy$ observed for pions, kaons, and protons, particularly observed at lower RHIC energies. While hydrodynamic models can reproduce this behavior for baryons, they fail to comprehensively describe the full range of experimental observations. To achieve a deeper understanding of the data, it is essential to develop advanced models that incorporate a combination of transport mechanisms, mean-field effects, and electromagnetic field interactions. Our study aims to help explore the interplay between particle production processes, baryon transport, and directed flow in heavy-ion collisions.

\section{Acknowledgments}
Financial assistance from Chinese Academy of Sciences (Grant No. XDB34000000) is gratefully acknowledged. MFT acknowledges support from ANSO Scholarship for Young Talents. We would like to thank our colleagues at the Institute of Modern Physics for many insightful discussions. Authors would like to acknowledge discussions with Sandeep Chatterjee and Tribhuban Parida during this work.

\bibliographystyle{apsrev4-1}
\bibliography{reference}

\begin{thebibliography}{45}%
\makeatletter
\providecommand \@ifxundefined [1]{%
 \@ifx{#1\undefined}
}%
\providecommand \@ifnum [1]{%
 \ifnum #1\expandafter \@firstoftwo
 \else \expandafter \@secondoftwo
 \fi
}%
\providecommand \@ifx [1]{%
 \ifx #1\expandafter \@firstoftwo
 \else \expandafter \@secondoftwo
 \fi
}%
\providecommand \natexlab [1]{#1}%
\providecommand \enquote  [1]{``#1''}%
\providecommand \bibnamefont  [1]{#1}%
\providecommand \bibfnamefont [1]{#1}%
\providecommand \citenamefont [1]{#1}%
\providecommand \href@noop [0]{\@secondoftwo}%
\providecommand \href [0]{\begingroup \@sanitize@url \@href}%
\providecommand \@href[1]{\@@startlink{#1}\@@href}%
\providecommand \@@href[1]{\endgroup#1\@@endlink}%
\providecommand \@sanitize@url [0]{\catcode `\\12\catcode `\$12\catcode
  `\&12\catcode `\#12\catcode `\^12\catcode `\_12\catcode `\%12\relax}%
\providecommand \@@startlink[1]{}%
\providecommand \@@endlink[0]{}%
\providecommand \url  [0]{\begingroup\@sanitize@url \@url }%
\providecommand \@url [1]{\endgroup\@href {#1}{\urlprefix }}%
\providecommand \urlprefix  [0]{URL }%
\providecommand \Eprint [0]{\href }%
\providecommand \doibase [0]{http://dx.doi.org/}%
\providecommand \selectlanguage [0]{\@gobble}%
\providecommand \bibinfo  [0]{\@secondoftwo}%
\providecommand \bibfield  [0]{\@secondoftwo}%
\providecommand \translation [1]{[#1]}%
\providecommand \BibitemOpen [0]{}%
\providecommand \bibitemStop [0]{}%
\providecommand \bibitemNoStop [0]{.\EOS\space}%
\providecommand \EOS [0]{\spacefactor3000\relax}%
\providecommand \BibitemShut  [1]{\csname bibitem#1\endcsname}%
\let\auto@bib@innerbib\@empty
\bibitem [{\citenamefont {Abelev}\ \emph {et~al.}(2008)\citenamefont {Abelev}
  \emph {et~al.}}]{STAR:2008jgm}%
  \BibitemOpen
  \bibfield  {author} {\bibinfo {author} {\bibfnamefont {B.~I.}\ \bibnamefont
  {Abelev}} \emph {et~al.} (\bibinfo {collaboration} {STAR}),\ }\href {\doibase
  10.1103/PhysRevLett.101.252301} {\bibfield  {journal} {\bibinfo  {journal}
  {Phys. Rev. Lett.}\ }\textbf {\bibinfo {volume} {101}},\ \bibinfo {pages}
  {252301} (\bibinfo {year} {2008})},\ \Eprint {http://arxiv.org/abs/0807.1518}
  {arXiv:0807.1518 [nucl-ex]} \BibitemShut {NoStop}%
\bibitem [{\citenamefont {Taseer}\ \emph {et~al.}(2024)\citenamefont {Taseer}
  \emph {et~al.}}]{Taseer:SQM2024talk}%
  \BibitemOpen
  \bibfield  {author} {\bibinfo {author} {\bibfnamefont {M.~F.}\ \bibnamefont
  {Taseer}} \emph {et~al.} (\bibinfo {collaboration} {for STAR
  Collaboration}),\ }\href {\doibase
  https://indico.in2p3.fr/event/29792/contributions/137162/} {\bibfield
  {journal} {\bibinfo  {journal} {Contribution to SQM 2024}\ } (\bibinfo {year}
  {2024}),\
  https://indico.in2p3.fr/event/29792/contributions/137162/}\BibitemShut
  {NoStop}%
\bibitem [{\citenamefont {Adams}\ \emph {et~al.}(2005)\citenamefont {Adams}
  \emph {et~al.}}]{Adams:2005dq}%
  \BibitemOpen
  \bibfield  {author} {\bibinfo {author} {\bibfnamefont {J.}~\bibnamefont
  {Adams}} \emph {et~al.} (\bibinfo {collaboration} {STAR}),\ }\href {\doibase
  10.1016/j.nuclphysa.2005.03.085} {\bibfield  {journal} {\bibinfo  {journal}
  {Nucl. Phys. A}\ }\textbf {\bibinfo {volume} {757}},\ \bibinfo {pages} {102}
  (\bibinfo {year} {2005})},\ \Eprint {http://arxiv.org/abs/nucl-ex/0501009}
  {arXiv:nucl-ex/0501009} \BibitemShut {NoStop}%
\bibitem [{\citenamefont {Adcox}\ \emph {et~al.}(2005)\citenamefont {Adcox}
  \emph {et~al.}}]{Adcox:2004mh}%
  \BibitemOpen
  \bibfield  {author} {\bibinfo {author} {\bibfnamefont {K.}~\bibnamefont
  {Adcox}} \emph {et~al.} (\bibinfo {collaboration} {PHENIX}),\ }\href
  {\doibase 10.1016/j.nuclphysa.2005.03.086} {\bibfield  {journal} {\bibinfo
  {journal} {Nucl. Phys. A}\ }\textbf {\bibinfo {volume} {757}},\ \bibinfo
  {pages} {184} (\bibinfo {year} {2005})},\ \Eprint
  {http://arxiv.org/abs/nucl-ex/0410003} {arXiv:nucl-ex/0410003} \BibitemShut
  {NoStop}%
\bibitem [{\citenamefont {Arsene}\ \emph {et~al.}(2005)\citenamefont {Arsene}
  \emph {et~al.}}]{BRAHMS:2004adc}%
  \BibitemOpen
  \bibfield  {author} {\bibinfo {author} {\bibfnamefont {I.}~\bibnamefont
  {Arsene}} \emph {et~al.} (\bibinfo {collaboration} {BRAHMS}),\ }\href
  {\doibase 10.1016/j.nuclphysa.2005.02.130} {\bibfield  {journal} {\bibinfo
  {journal} {Nucl. Phys. A}\ }\textbf {\bibinfo {volume} {757}},\ \bibinfo
  {pages} {1} (\bibinfo {year} {2005})},\ \Eprint
  {http://arxiv.org/abs/nucl-ex/0410020} {arXiv:nucl-ex/0410020} \BibitemShut
  {NoStop}%
\bibitem [{\citenamefont {Harris}\ and\ \citenamefont
  {Muller}(1996)}]{Harris:1996zx}%
  \BibitemOpen
  \bibfield  {author} {\bibinfo {author} {\bibfnamefont {J.~W.}\ \bibnamefont
  {Harris}}\ and\ \bibinfo {author} {\bibfnamefont {B.}~\bibnamefont
  {Muller}},\ }\href {\doibase 10.1146/annurev.nucl.46.1.71} {\bibfield
  {journal} {\bibinfo  {journal} {Ann. Rev. Nucl. Part. Sci.}\ }\textbf
  {\bibinfo {volume} {46}},\ \bibinfo {pages} {71} (\bibinfo {year} {1996})},\
  \Eprint {http://arxiv.org/abs/hep-ph/9602235} {arXiv:hep-ph/9602235}
  \BibitemShut {NoStop}%
\bibitem [{\citenamefont {Muller}\ \emph {et~al.}(2012)\citenamefont {Muller},
  \citenamefont {Schukraft},\ and\ \citenamefont {Wyslouch}}]{Muller:2012zq}%
  \BibitemOpen
  \bibfield  {author} {\bibinfo {author} {\bibfnamefont {B.}~\bibnamefont
  {Muller}}, \bibinfo {author} {\bibfnamefont {J.}~\bibnamefont {Schukraft}}, \
  and\ \bibinfo {author} {\bibfnamefont {B.}~\bibnamefont {Wyslouch}},\ }\href
  {\doibase 10.1146/annurev-nucl-102711-094910} {\bibfield  {journal} {\bibinfo
   {journal} {Ann. Rev. Nucl. Part. Sci.}\ }\textbf {\bibinfo {volume} {62}},\
  \bibinfo {pages} {361} (\bibinfo {year} {2012})},\ \Eprint
  {http://arxiv.org/abs/1202.3233} {arXiv:1202.3233 [hep-ex]} \BibitemShut
  {NoStop}%
\bibitem [{\citenamefont {Braun-Munzinger}\ and\ \citenamefont
  {Stachel}(2007)}]{Braun-Munzinger:2007edi}%
  \BibitemOpen
  \bibfield  {author} {\bibinfo {author} {\bibfnamefont {P.}~\bibnamefont
  {Braun-Munzinger}}\ and\ \bibinfo {author} {\bibfnamefont {J.}~\bibnamefont
  {Stachel}},\ }\href {\doibase 10.1038/nature06080} {\bibfield  {journal}
  {\bibinfo  {journal} {Nature}\ }\textbf {\bibinfo {volume} {448}},\ \bibinfo
  {pages} {302} (\bibinfo {year} {2007})}\BibitemShut {NoStop}%
\bibitem [{\citenamefont {Ollitrault}(1992)}]{Ollitrault:1992bk}%
  \BibitemOpen
  \bibfield  {author} {\bibinfo {author} {\bibfnamefont {J.-Y.}\ \bibnamefont
  {Ollitrault}},\ }\href {\doibase 10.1103/PhysRevD.46.229} {\bibfield
  {journal} {\bibinfo  {journal} {Phys. Rev. D}\ }\textbf {\bibinfo {volume}
  {46}},\ \bibinfo {pages} {229} (\bibinfo {year} {1992})}\BibitemShut
  {NoStop}%
\bibitem [{\citenamefont {Poskanzer}\ and\ \citenamefont
  {Voloshin}(1998)}]{Poskanzer:1998yz}%
  \BibitemOpen
  \bibfield  {author} {\bibinfo {author} {\bibfnamefont {A.~M.}\ \bibnamefont
  {Poskanzer}}\ and\ \bibinfo {author} {\bibfnamefont {S.~A.}\ \bibnamefont
  {Voloshin}},\ }\href {\doibase 10.1103/PhysRevC.58.1671} {\bibfield
  {journal} {\bibinfo  {journal} {Phys. Rev. C}\ }\textbf {\bibinfo {volume}
  {58}},\ \bibinfo {pages} {1671} (\bibinfo {year} {1998})},\ \Eprint
  {http://arxiv.org/abs/nucl-ex/9805001} {arXiv:nucl-ex/9805001} \BibitemShut
  {NoStop}%
\bibitem [{\citenamefont {Voloshin}\ and\ \citenamefont
  {Zhang}(1996)}]{Voloshin:1994mz}%
  \BibitemOpen
  \bibfield  {author} {\bibinfo {author} {\bibfnamefont {S.}~\bibnamefont
  {Voloshin}}\ and\ \bibinfo {author} {\bibfnamefont {Y.}~\bibnamefont
  {Zhang}},\ }\href {\doibase 10.1007/s002880050141} {\bibfield  {journal}
  {\bibinfo  {journal} {Z. Phys. C}\ }\textbf {\bibinfo {volume} {70}},\
  \bibinfo {pages} {665} (\bibinfo {year} {1996})},\ \Eprint
  {http://arxiv.org/abs/hep-ph/9407282} {arXiv:hep-ph/9407282} \BibitemShut
  {NoStop}%
\bibitem [{\citenamefont {Brachmann}\ \emph {et~al.}(2000)\citenamefont
  {Brachmann}, \citenamefont {Soff}, \citenamefont {Dumitru}, \citenamefont
  {Stoecker}, \citenamefont {Maruhn}, \citenamefont {Greiner}, \citenamefont
  {Bravina},\ and\ \citenamefont {Rischke}}]{Brachmann:1999xt}%
  \BibitemOpen
  \bibfield  {author} {\bibinfo {author} {\bibfnamefont {J.}~\bibnamefont
  {Brachmann}}, \bibinfo {author} {\bibfnamefont {S.}~\bibnamefont {Soff}},
  \bibinfo {author} {\bibfnamefont {A.}~\bibnamefont {Dumitru}}, \bibinfo
  {author} {\bibfnamefont {H.}~\bibnamefont {Stoecker}}, \bibinfo {author}
  {\bibfnamefont {J.~A.}\ \bibnamefont {Maruhn}}, \bibinfo {author}
  {\bibfnamefont {W.}~\bibnamefont {Greiner}}, \bibinfo {author} {\bibfnamefont
  {L.~V.}\ \bibnamefont {Bravina}}, \ and\ \bibinfo {author} {\bibfnamefont
  {D.~H.}\ \bibnamefont {Rischke}},\ }\href {\doibase
  10.1103/PhysRevC.61.024909} {\bibfield  {journal} {\bibinfo  {journal} {Phys.
  Rev. C}\ }\textbf {\bibinfo {volume} {61}},\ \bibinfo {pages} {024909}
  (\bibinfo {year} {2000})},\ \Eprint {http://arxiv.org/abs/nucl-th/9908010}
  {arXiv:nucl-th/9908010} \BibitemShut {NoStop}%
\bibitem [{\citenamefont {Chung}\ \emph {et~al.}(2000)\citenamefont {Chung}
  \emph {et~al.}}]{E895:2000sor}%
  \BibitemOpen
  \bibfield  {author} {\bibinfo {author} {\bibfnamefont {P.}~\bibnamefont
  {Chung}} \emph {et~al.} (\bibinfo {collaboration} {E895}),\ }\href {\doibase
  10.1103/PhysRevLett.85.940} {\bibfield  {journal} {\bibinfo  {journal} {Phys.
  Rev. Lett.}\ }\textbf {\bibinfo {volume} {85}},\ \bibinfo {pages} {940}
  (\bibinfo {year} {2000})},\ \Eprint {http://arxiv.org/abs/nucl-ex/0101003}
  {arXiv:nucl-ex/0101003} \BibitemShut {NoStop}%
\bibitem [{\citenamefont {Chung}\ \emph {et~al.}(2001)\citenamefont {Chung}
  \emph {et~al.}}]{Chung:2001je}%
  \BibitemOpen
  \bibfield  {author} {\bibinfo {author} {\bibfnamefont {P.}~\bibnamefont
  {Chung}} \emph {et~al.},\ }\href {\doibase 10.1103/PhysRevLett.86.2533}
  {\bibfield  {journal} {\bibinfo  {journal} {Phys. Rev. Lett.}\ }\textbf
  {\bibinfo {volume} {86}},\ \bibinfo {pages} {2533} (\bibinfo {year}
  {2001})},\ \Eprint {http://arxiv.org/abs/nucl-ex/0101002}
  {arXiv:nucl-ex/0101002} \BibitemShut {NoStop}%
\bibitem [{\citenamefont {Alt}\ \emph {et~al.}(2003)\citenamefont {Alt} \emph
  {et~al.}}]{NA49:2003njx}%
  \BibitemOpen
  \bibfield  {author} {\bibinfo {author} {\bibfnamefont {C.}~\bibnamefont
  {Alt}} \emph {et~al.} (\bibinfo {collaboration} {NA49}),\ }\href {\doibase
  10.1103/PhysRevC.68.034903} {\bibfield  {journal} {\bibinfo  {journal} {Phys.
  Rev. C}\ }\textbf {\bibinfo {volume} {68}},\ \bibinfo {pages} {034903}
  (\bibinfo {year} {2003})},\ \Eprint {http://arxiv.org/abs/nucl-ex/0303001}
  {arXiv:nucl-ex/0303001} \BibitemShut {NoStop}%
\bibitem [{\citenamefont {Adams}\ \emph {et~al.}(2004)\citenamefont {Adams}
  \emph {et~al.}}]{STAR:2003xyj}%
  \BibitemOpen
  \bibfield  {author} {\bibinfo {author} {\bibfnamefont {J.}~\bibnamefont
  {Adams}} \emph {et~al.} (\bibinfo {collaboration} {STAR}),\ }\href {\doibase
  10.1103/PhysRevLett.127.069901} {\bibfield  {journal} {\bibinfo  {journal}
  {Phys. Rev. Lett.}\ }\textbf {\bibinfo {volume} {92}},\ \bibinfo {pages}
  {062301} (\bibinfo {year} {2004})},\ \bibinfo {note} {[Erratum:
  Phys.Rev.Lett. 127, 069901 (2021)]},\ \Eprint
  {http://arxiv.org/abs/nucl-ex/0310029} {arXiv:nucl-ex/0310029} \BibitemShut
  {NoStop}%
\bibitem [{\citenamefont {Back}\ \emph {et~al.}(2006)\citenamefont {Back} \emph
  {et~al.}}]{PHOBOS:2005ylx}%
  \BibitemOpen
  \bibfield  {author} {\bibinfo {author} {\bibfnamefont {B.~B.}\ \bibnamefont
  {Back}} \emph {et~al.} (\bibinfo {collaboration} {PHOBOS}),\ }\href {\doibase
  10.1103/PhysRevLett.97.012301} {\bibfield  {journal} {\bibinfo  {journal}
  {Phys. Rev. Lett.}\ }\textbf {\bibinfo {volume} {97}},\ \bibinfo {pages}
  {012301} (\bibinfo {year} {2006})},\ \Eprint
  {http://arxiv.org/abs/nucl-ex/0511045} {arXiv:nucl-ex/0511045} \BibitemShut
  {NoStop}%
\bibitem [{\citenamefont {Abelev}\ \emph {et~al.}(2013)\citenamefont {Abelev}
  \emph {et~al.}}]{ALICE:2013xri}%
  \BibitemOpen
  \bibfield  {author} {\bibinfo {author} {\bibfnamefont {B.}~\bibnamefont
  {Abelev}} \emph {et~al.} (\bibinfo {collaboration} {ALICE}),\ }\href
  {\doibase 10.1103/PhysRevLett.111.232302} {\bibfield  {journal} {\bibinfo
  {journal} {Phys. Rev. Lett.}\ }\textbf {\bibinfo {volume} {111}},\ \bibinfo
  {pages} {232302} (\bibinfo {year} {2013})},\ \Eprint
  {http://arxiv.org/abs/1306.4145} {arXiv:1306.4145 [nucl-ex]} \BibitemShut
  {NoStop}%
\bibitem [{\citenamefont {Stoecker}(2005)}]{Stoecker:2004qu}%
  \BibitemOpen
  \bibfield  {author} {\bibinfo {author} {\bibfnamefont {H.}~\bibnamefont
  {Stoecker}},\ }\href {\doibase 10.1016/j.nuclphysa.2004.12.074} {\bibfield
  {journal} {\bibinfo  {journal} {Nucl. Phys. A}\ }\textbf {\bibinfo {volume}
  {750}},\ \bibinfo {pages} {121} (\bibinfo {year} {2005})},\ \Eprint
  {http://arxiv.org/abs/nucl-th/0406018} {arXiv:nucl-th/0406018} \BibitemShut
  {NoStop}%
\bibitem [{\citenamefont {Nara}\ \emph {et~al.}(2016)\citenamefont {Nara},
  \citenamefont {Niemi}, \citenamefont {Ohnishi},\ and\ \citenamefont
  {St\"ocker}}]{Nara:2016phs}%
  \BibitemOpen
  \bibfield  {author} {\bibinfo {author} {\bibfnamefont {Y.}~\bibnamefont
  {Nara}}, \bibinfo {author} {\bibfnamefont {H.}~\bibnamefont {Niemi}},
  \bibinfo {author} {\bibfnamefont {A.}~\bibnamefont {Ohnishi}}, \ and\
  \bibinfo {author} {\bibfnamefont {H.}~\bibnamefont {St\"ocker}},\ }\href
  {\doibase 10.1103/PhysRevC.94.034906} {\bibfield  {journal} {\bibinfo
  {journal} {Phys. Rev. C}\ }\textbf {\bibinfo {volume} {94}},\ \bibinfo
  {pages} {034906} (\bibinfo {year} {2016})},\ \Eprint
  {http://arxiv.org/abs/1601.07692} {arXiv:1601.07692 [hep-ph]} \BibitemShut
  {NoStop}%
\bibitem [{\citenamefont {Konchakovski}\ \emph {et~al.}(2014)\citenamefont
  {Konchakovski}, \citenamefont {Cassing}, \citenamefont {Ivanov},\ and\
  \citenamefont {Toneev}}]{Konchakovski:2014gda}%
  \BibitemOpen
  \bibfield  {author} {\bibinfo {author} {\bibfnamefont {V.~P.}\ \bibnamefont
  {Konchakovski}}, \bibinfo {author} {\bibfnamefont {W.}~\bibnamefont
  {Cassing}}, \bibinfo {author} {\bibfnamefont {Y.~B.}\ \bibnamefont {Ivanov}},
  \ and\ \bibinfo {author} {\bibfnamefont {V.~D.}\ \bibnamefont {Toneev}},\
  }\href {\doibase 10.1103/PhysRevC.90.014903} {\bibfield  {journal} {\bibinfo
  {journal} {Phys. Rev. C}\ }\textbf {\bibinfo {volume} {90}},\ \bibinfo
  {pages} {014903} (\bibinfo {year} {2014})},\ \Eprint
  {http://arxiv.org/abs/1404.2765} {arXiv:1404.2765 [nucl-th]} \BibitemShut
  {NoStop}%
\bibitem [{\citenamefont {Adamczyk}\ \emph {et~al.}(2014)\citenamefont
  {Adamczyk} \emph {et~al.}}]{STAR:2014clz}%
  \BibitemOpen
  \bibfield  {author} {\bibinfo {author} {\bibfnamefont {L.}~\bibnamefont
  {Adamczyk}} \emph {et~al.} (\bibinfo {collaboration} {STAR}),\ }\href
  {\doibase 10.1103/PhysRevLett.112.162301} {\bibfield  {journal} {\bibinfo
  {journal} {Phys. Rev. Lett.}\ }\textbf {\bibinfo {volume} {112}},\ \bibinfo
  {pages} {162301} (\bibinfo {year} {2014})},\ \Eprint
  {http://arxiv.org/abs/1401.3043} {arXiv:1401.3043 [nucl-ex]} \BibitemShut
  {NoStop}%
\bibitem [{\citenamefont {Adamczyk}\ \emph {et~al.}(2018)\citenamefont
  {Adamczyk} \emph {et~al.}}]{STAR:2017okv}%
  \BibitemOpen
  \bibfield  {author} {\bibinfo {author} {\bibfnamefont {L.}~\bibnamefont
  {Adamczyk}} \emph {et~al.} (\bibinfo {collaboration} {STAR}),\ }\href
  {\doibase 10.1103/PhysRevLett.120.062301} {\bibfield  {journal} {\bibinfo
  {journal} {Phys. Rev. Lett.}\ }\textbf {\bibinfo {volume} {120}},\ \bibinfo
  {pages} {062301} (\bibinfo {year} {2018})},\ \Eprint
  {http://arxiv.org/abs/1708.07132} {arXiv:1708.07132 [hep-ex]} \BibitemShut
  {NoStop}%
\bibitem [{\citenamefont {Singha}\ \emph {et~al.}(2016)\citenamefont {Singha},
  \citenamefont {Shanmuganathan},\ and\ \citenamefont
  {Keane}}]{Singha:2016mna}%
  \BibitemOpen
  \bibfield  {author} {\bibinfo {author} {\bibfnamefont {S.}~\bibnamefont
  {Singha}}, \bibinfo {author} {\bibfnamefont {P.}~\bibnamefont
  {Shanmuganathan}}, \ and\ \bibinfo {author} {\bibfnamefont {D.}~\bibnamefont
  {Keane}},\ }\href {\doibase 10.1155/2016/2836989} {\bibfield  {journal}
  {\bibinfo  {journal} {Adv. High Energy Phys.}\ }\textbf {\bibinfo {volume}
  {2016}},\ \bibinfo {pages} {2836989} (\bibinfo {year} {2016})},\ \Eprint
  {http://arxiv.org/abs/1610.00646} {arXiv:1610.00646 [nucl-ex]} \BibitemShut
  {NoStop}%
\bibitem [{\citenamefont {Kharzeev}\ \emph {et~al.}(2008)\citenamefont
  {Kharzeev}, \citenamefont {McLerran},\ and\ \citenamefont
  {Warringa}}]{Kharzeev:2007jp}%
  \BibitemOpen
  \bibfield  {author} {\bibinfo {author} {\bibfnamefont {D.~E.}\ \bibnamefont
  {Kharzeev}}, \bibinfo {author} {\bibfnamefont {L.~D.}\ \bibnamefont
  {McLerran}}, \ and\ \bibinfo {author} {\bibfnamefont {H.~J.}\ \bibnamefont
  {Warringa}},\ }\href {\doibase 10.1016/j.nuclphysa.2008.02.298} {\bibfield
  {journal} {\bibinfo  {journal} {Nucl. Phys. A}\ }\textbf {\bibinfo {volume}
  {803}},\ \bibinfo {pages} {227} (\bibinfo {year} {2008})},\ \Eprint
  {http://arxiv.org/abs/0711.0950} {arXiv:0711.0950 [hep-ph]} \BibitemShut
  {NoStop}%
\bibitem [{\citenamefont {McLerran}\ and\ \citenamefont
  {Skokov}(2014)}]{McLerran:2013hla}%
  \BibitemOpen
  \bibfield  {author} {\bibinfo {author} {\bibfnamefont {L.}~\bibnamefont
  {McLerran}}\ and\ \bibinfo {author} {\bibfnamefont {V.}~\bibnamefont
  {Skokov}},\ }\href {\doibase 10.1016/j.nuclphysa.2014.05.008} {\bibfield
  {journal} {\bibinfo  {journal} {Nucl. Phys. A}\ }\textbf {\bibinfo {volume}
  {929}},\ \bibinfo {pages} {184} (\bibinfo {year} {2014})},\ \Eprint
  {http://arxiv.org/abs/1305.0774} {arXiv:1305.0774 [hep-ph]} \BibitemShut
  {NoStop}%
\bibitem [{\citenamefont {Gursoy}\ \emph {et~al.}(2014)\citenamefont {Gursoy},
  \citenamefont {Kharzeev},\ and\ \citenamefont {Rajagopal}}]{Gursoy:2014aka}%
  \BibitemOpen
  \bibfield  {author} {\bibinfo {author} {\bibfnamefont {U.}~\bibnamefont
  {Gursoy}}, \bibinfo {author} {\bibfnamefont {D.}~\bibnamefont {Kharzeev}}, \
  and\ \bibinfo {author} {\bibfnamefont {K.}~\bibnamefont {Rajagopal}},\ }\href
  {\doibase 10.1103/PhysRevC.89.054905} {\bibfield  {journal} {\bibinfo
  {journal} {Phys. Rev. C}\ }\textbf {\bibinfo {volume} {89}},\ \bibinfo
  {pages} {054905} (\bibinfo {year} {2014})},\ \Eprint
  {http://arxiv.org/abs/1401.3805} {arXiv:1401.3805 [hep-ph]} \BibitemShut
  {NoStop}%
\bibitem [{\citenamefont {G\"ursoy}\ \emph {et~al.}(2018)\citenamefont
  {G\"ursoy}, \citenamefont {Kharzeev}, \citenamefont {Marcus}, \citenamefont
  {Rajagopal},\ and\ \citenamefont {Shen}}]{Gursoy:2018yai}%
  \BibitemOpen
  \bibfield  {author} {\bibinfo {author} {\bibfnamefont {U.}~\bibnamefont
  {G\"ursoy}}, \bibinfo {author} {\bibfnamefont {D.}~\bibnamefont {Kharzeev}},
  \bibinfo {author} {\bibfnamefont {E.}~\bibnamefont {Marcus}}, \bibinfo
  {author} {\bibfnamefont {K.}~\bibnamefont {Rajagopal}}, \ and\ \bibinfo
  {author} {\bibfnamefont {C.}~\bibnamefont {Shen}},\ }\href {\doibase
  10.1103/PhysRevC.98.055201} {\bibfield  {journal} {\bibinfo  {journal} {Phys.
  Rev. C}\ }\textbf {\bibinfo {volume} {98}},\ \bibinfo {pages} {055201}
  (\bibinfo {year} {2018})},\ \Eprint {http://arxiv.org/abs/1806.05288}
  {arXiv:1806.05288 [hep-ph]} \BibitemShut {NoStop}%
\bibitem [{\citenamefont {Abdulhamid}\ \emph {et~al.}(2024)\citenamefont
  {Abdulhamid} \emph {et~al.}}]{STAR:2023jdd}%
  \BibitemOpen
  \bibfield  {author} {\bibinfo {author} {\bibfnamefont {M.~I.}\ \bibnamefont
  {Abdulhamid}} \emph {et~al.} (\bibinfo {collaboration} {STAR}),\ }\href
  {\doibase 10.1103/PhysRevX.14.011028} {\bibfield  {journal} {\bibinfo
  {journal} {Phys. Rev. X}\ }\textbf {\bibinfo {volume} {14}},\ \bibinfo
  {pages} {011028} (\bibinfo {year} {2024})},\ \Eprint
  {http://arxiv.org/abs/2304.03430} {arXiv:2304.03430 [nucl-ex]} \BibitemShut
  {NoStop}%
\bibitem [{\citenamefont {STAR}(2023)}]{STAR:2023wjl}%
  \BibitemOpen
  \bibfield  {author} {\bibinfo {author} {\bibnamefont {STAR}} (\bibinfo
  {collaboration} {STAR}),\ }\href@noop {} {\  (\bibinfo {year} {2023})},\
  \Eprint {http://arxiv.org/abs/2304.02831} {arXiv:2304.02831 [nucl-ex]}
  \BibitemShut {NoStop}%
\bibitem [{\citenamefont {Parida}\ and\ \citenamefont
  {Chatterjee}(2023{\natexlab{a}})}]{Parida:2023ldu}%
  \BibitemOpen
  \bibfield  {author} {\bibinfo {author} {\bibfnamefont {T.}~\bibnamefont
  {Parida}}\ and\ \bibinfo {author} {\bibfnamefont {S.}~\bibnamefont
  {Chatterjee}},\ }\href@noop {} {\  (\bibinfo {year} {2023}{\natexlab{a}})},\
  \Eprint {http://arxiv.org/abs/2305.08806} {arXiv:2305.08806 [nucl-th]}
  \BibitemShut {NoStop}%
\bibitem [{\citenamefont {Parida}\ and\ \citenamefont
  {Chatterjee}(2023{\natexlab{b}})}]{Parida:2023rux}%
  \BibitemOpen
  \bibfield  {author} {\bibinfo {author} {\bibfnamefont {T.}~\bibnamefont
  {Parida}}\ and\ \bibinfo {author} {\bibfnamefont {S.}~\bibnamefont
  {Chatterjee}},\ }\href@noop {} {\  (\bibinfo {year} {2023}{\natexlab{b}})},\
  \Eprint {http://arxiv.org/abs/2305.10371} {arXiv:2305.10371 [nucl-th]}
  \BibitemShut {NoStop}%
\bibitem [{\citenamefont {Parida}\ \emph {et~al.}(2025)\citenamefont {Parida},
  \citenamefont {Chatterjee},\ and\ \citenamefont {Singha}}]{Parida:2025ddt}%
  \BibitemOpen
  \bibfield  {author} {\bibinfo {author} {\bibfnamefont {T.}~\bibnamefont
  {Parida}}, \bibinfo {author} {\bibfnamefont {S.}~\bibnamefont {Chatterjee}},
  \ and\ \bibinfo {author} {\bibfnamefont {S.}~\bibnamefont {Singha}},\
  }\href@noop {} {\  (\bibinfo {year} {2025})},\ \Eprint
  {http://arxiv.org/abs/2503.04660} {arXiv:2503.04660 [nucl-th]} \BibitemShut
  {NoStop}%
\bibitem [{\citenamefont {Isse}\ \emph {et~al.}(2005)\citenamefont {Isse},
  \citenamefont {Ohnishi}, \citenamefont {Otuka}, \citenamefont {Sahu},\ and\
  \citenamefont {Nara}}]{Isse:2005nk}%
  \BibitemOpen
  \bibfield  {author} {\bibinfo {author} {\bibfnamefont {M.}~\bibnamefont
  {Isse}}, \bibinfo {author} {\bibfnamefont {A.}~\bibnamefont {Ohnishi}},
  \bibinfo {author} {\bibfnamefont {N.}~\bibnamefont {Otuka}}, \bibinfo
  {author} {\bibfnamefont {P.~K.}\ \bibnamefont {Sahu}}, \ and\ \bibinfo
  {author} {\bibfnamefont {Y.}~\bibnamefont {Nara}},\ }\href {\doibase
  10.1103/PhysRevC.72.064908} {\bibfield  {journal} {\bibinfo  {journal} {Phys.
  Rev. C}\ }\textbf {\bibinfo {volume} {72}},\ \bibinfo {pages} {064908}
  (\bibinfo {year} {2005})},\ \Eprint {http://arxiv.org/abs/nucl-th/0502058}
  {arXiv:nucl-th/0502058} \BibitemShut {NoStop}%
\bibitem [{\citenamefont {Nara}\ and\ \citenamefont
  {Ohnishi}(2022)}]{Nara:2021fuu}%
  \BibitemOpen
  \bibfield  {author} {\bibinfo {author} {\bibfnamefont {Y.}~\bibnamefont
  {Nara}}\ and\ \bibinfo {author} {\bibfnamefont {A.}~\bibnamefont {Ohnishi}},\
  }\href {\doibase 10.1103/PhysRevC.105.014911} {\bibfield  {journal} {\bibinfo
   {journal} {Phys. Rev. C}\ }\textbf {\bibinfo {volume} {105}},\ \bibinfo
  {pages} {014911} (\bibinfo {year} {2022})},\ \Eprint
  {http://arxiv.org/abs/2109.07594} {arXiv:2109.07594 [nucl-th]} \BibitemShut
  {NoStop}%
\bibitem [{\citenamefont {Nara}\ \emph {et~al.}(2020)\citenamefont {Nara},
  \citenamefont {Maruyama},\ and\ \citenamefont {Stoecker}}]{Nara:2020ztb}%
  \BibitemOpen
  \bibfield  {author} {\bibinfo {author} {\bibfnamefont {Y.}~\bibnamefont
  {Nara}}, \bibinfo {author} {\bibfnamefont {T.}~\bibnamefont {Maruyama}}, \
  and\ \bibinfo {author} {\bibfnamefont {H.}~\bibnamefont {Stoecker}},\ }\href
  {\doibase 10.1103/PhysRevC.102.024913} {\bibfield  {journal} {\bibinfo
  {journal} {Phys. Rev. C}\ }\textbf {\bibinfo {volume} {102}},\ \bibinfo
  {pages} {024913} (\bibinfo {year} {2020})},\ \Eprint
  {http://arxiv.org/abs/2004.05550} {arXiv:2004.05550 [nucl-th]} \BibitemShut
  {NoStop}%
\bibitem [{\citenamefont {Lin}\ \emph {et~al.}(2005)\citenamefont {Lin},
  \citenamefont {Ko}, \citenamefont {Li}, \citenamefont {Zhang},\ and\
  \citenamefont {Pal}}]{Lin:2004en}%
  \BibitemOpen
  \bibfield  {author} {\bibinfo {author} {\bibfnamefont {Z.-W.}\ \bibnamefont
  {Lin}}, \bibinfo {author} {\bibfnamefont {C.~M.}\ \bibnamefont {Ko}},
  \bibinfo {author} {\bibfnamefont {B.-A.}\ \bibnamefont {Li}}, \bibinfo
  {author} {\bibfnamefont {B.}~\bibnamefont {Zhang}}, \ and\ \bibinfo {author}
  {\bibfnamefont {S.}~\bibnamefont {Pal}},\ }\href {\doibase
  10.1103/PhysRevC.72.064901} {\bibfield  {journal} {\bibinfo  {journal} {Phys.
  Rev. C}\ }\textbf {\bibinfo {volume} {72}},\ \bibinfo {pages} {064901}
  (\bibinfo {year} {2005})},\ \Eprint {http://arxiv.org/abs/nucl-th/0411110}
  {arXiv:nucl-th/0411110} \BibitemShut {NoStop}%
\bibitem [{\citenamefont {Adamczyk}\ \emph {et~al.}(2017)\citenamefont
  {Adamczyk} \emph {et~al.}}]{STAR:2017sal}%
  \BibitemOpen
  \bibfield  {author} {\bibinfo {author} {\bibfnamefont {L.}~\bibnamefont
  {Adamczyk}} \emph {et~al.} (\bibinfo {collaboration} {STAR}),\ }\href
  {\doibase 10.1103/PhysRevC.96.044904} {\bibfield  {journal} {\bibinfo
  {journal} {Phys. Rev. C}\ }\textbf {\bibinfo {volume} {96}},\ \bibinfo
  {pages} {044904} (\bibinfo {year} {2017})},\ \Eprint
  {http://arxiv.org/abs/1701.07065} {arXiv:1701.07065 [nucl-ex]} \BibitemShut
  {NoStop}%
\bibitem [{\citenamefont {Wang}\ and\ \citenamefont
  {Gyulassy}(1991)}]{Wang:1991hta}%
  \BibitemOpen
  \bibfield  {author} {\bibinfo {author} {\bibfnamefont {X.-N.}\ \bibnamefont
  {Wang}}\ and\ \bibinfo {author} {\bibfnamefont {M.}~\bibnamefont
  {Gyulassy}},\ }\href {\doibase 10.1103/PhysRevD.44.3501} {\bibfield
  {journal} {\bibinfo  {journal} {Phys. Rev. D}\ }\textbf {\bibinfo {volume}
  {44}},\ \bibinfo {pages} {3501} (\bibinfo {year} {1991})}\BibitemShut
  {NoStop}%
\bibitem [{\citenamefont {Zhang}(1998)}]{Zhang:1997ej}%
  \BibitemOpen
  \bibfield  {author} {\bibinfo {author} {\bibfnamefont {B.}~\bibnamefont
  {Zhang}},\ }\href {\doibase 10.1016/S0010-4655(98)00010-1} {\bibfield
  {journal} {\bibinfo  {journal} {Comput. Phys. Commun.}\ }\textbf {\bibinfo
  {volume} {109}},\ \bibinfo {pages} {193} (\bibinfo {year} {1998})},\ \Eprint
  {http://arxiv.org/abs/nucl-th/9709009} {arXiv:nucl-th/9709009} \BibitemShut
  {NoStop}%
\bibitem [{\citenamefont {Andersson}\ \emph
  {et~al.}(1983{\natexlab{a}})\citenamefont {Andersson}, \citenamefont
  {Gustafson},\ and\ \citenamefont {Soderberg}}]{Andersson:1983jt}%
  \BibitemOpen
  \bibfield  {author} {\bibinfo {author} {\bibfnamefont {B.}~\bibnamefont
  {Andersson}}, \bibinfo {author} {\bibfnamefont {G.}~\bibnamefont
  {Gustafson}}, \ and\ \bibinfo {author} {\bibfnamefont {B.}~\bibnamefont
  {Soderberg}},\ }\href {\doibase 10.1007/BF01407824} {\bibfield  {journal}
  {\bibinfo  {journal} {Z. Phys. C}\ }\textbf {\bibinfo {volume} {20}},\
  \bibinfo {pages} {317} (\bibinfo {year} {1983}{\natexlab{a}})}\BibitemShut
  {NoStop}%
\bibitem [{\citenamefont {Andersson}\ \emph
  {et~al.}(1983{\natexlab{b}})\citenamefont {Andersson}, \citenamefont
  {Gustafson}, \citenamefont {Ingelman},\ and\ \citenamefont
  {Sjostrand}}]{Andersson:1983ia}%
  \BibitemOpen
  \bibfield  {author} {\bibinfo {author} {\bibfnamefont {B.}~\bibnamefont
  {Andersson}}, \bibinfo {author} {\bibfnamefont {G.}~\bibnamefont
  {Gustafson}}, \bibinfo {author} {\bibfnamefont {G.}~\bibnamefont {Ingelman}},
  \ and\ \bibinfo {author} {\bibfnamefont {T.}~\bibnamefont {Sjostrand}},\
  }\href {\doibase 10.1016/0370-1573(83)90080-7} {\bibfield  {journal}
  {\bibinfo  {journal} {Phys. Rept.}\ }\textbf {\bibinfo {volume} {97}},\
  \bibinfo {pages} {31} (\bibinfo {year} {1983}{\natexlab{b}})}\BibitemShut
  {NoStop}%
\bibitem [{\citenamefont {Sjostrand}(1994)}]{Sjostrand:1993yb}%
  \BibitemOpen
  \bibfield  {author} {\bibinfo {author} {\bibfnamefont {T.}~\bibnamefont
  {Sjostrand}},\ }\href {\doibase 10.1016/0010-4655(94)90132-5} {\bibfield
  {journal} {\bibinfo  {journal} {Comput. Phys. Commun.}\ }\textbf {\bibinfo
  {volume} {82}},\ \bibinfo {pages} {74} (\bibinfo {year} {1994})}\BibitemShut
  {NoStop}%
\bibitem [{\citenamefont {Li}\ and\ \citenamefont {Ko}(1995)}]{Li:1995pra}%
  \BibitemOpen
  \bibfield  {author} {\bibinfo {author} {\bibfnamefont {B.-A.}\ \bibnamefont
  {Li}}\ and\ \bibinfo {author} {\bibfnamefont {C.~M.}\ \bibnamefont {Ko}},\
  }\href {\doibase 10.1103/PhysRevC.52.2037} {\bibfield  {journal} {\bibinfo
  {journal} {Phys. Rev. C}\ }\textbf {\bibinfo {volume} {52}},\ \bibinfo
  {pages} {2037} (\bibinfo {year} {1995})},\ \Eprint
  {http://arxiv.org/abs/nucl-th/9505016} {arXiv:nucl-th/9505016} \BibitemShut
  {NoStop}%
\bibitem [{\citenamefont {Lin}\ \emph {et~al.}(2001)\citenamefont {Lin},
  \citenamefont {Pal}, \citenamefont {Ko}, \citenamefont {Li},\ and\
  \citenamefont {Zhang}}]{Lin:2000cx}%
  \BibitemOpen
  \bibfield  {author} {\bibinfo {author} {\bibfnamefont {Z.-w.}\ \bibnamefont
  {Lin}}, \bibinfo {author} {\bibfnamefont {S.}~\bibnamefont {Pal}}, \bibinfo
  {author} {\bibfnamefont {C.~M.}\ \bibnamefont {Ko}}, \bibinfo {author}
  {\bibfnamefont {B.-A.}\ \bibnamefont {Li}}, \ and\ \bibinfo {author}
  {\bibfnamefont {B.}~\bibnamefont {Zhang}},\ }\href {\doibase
  10.1103/PhysRevC.64.011902} {\bibfield  {journal} {\bibinfo  {journal} {Phys.
  Rev. C}\ }\textbf {\bibinfo {volume} {64}},\ \bibinfo {pages} {011902}
  (\bibinfo {year} {2001})},\ \Eprint {http://arxiv.org/abs/nucl-th/0011059}
  {arXiv:nucl-th/0011059} \BibitemShut {NoStop}%
\end{thebibliography}%
\end{document}